# Dose conversion coefficients for Chinese reference adult male and female voxel phantoms from idealized neutron exposures*

LIU Huan(刘欢)[1,2], LI Jun-Li(李君利)[1,2], QIU Rui(邱睿)[1,2;1)], YANG Yue(杨月)[1,2], PAN Yu-Xi(潘羽晞)[1,2], LIU Li-Ye(刘立业)[3]

[1]Department of Engineering Physics, Tsinghua University, Beijing 100084, China
[2]Key Laboratory of Particle & Radiation Imaging, Ministry of Education, Beijing 100084, China
[3]China Institute for Radiation Protection, Taiyuan 030006, China

**Abstract:** A new set of fluence-to-dose conversion coefficients based on the Chinese reference adult voxel phantoms CRAM and CRAF are presented for six idealized external neutron exposures from $10^{-8}$ MeV to 20 MeV. The voxel phantoms CRAM and CRAF were adjusted from the previous phantoms CNMAN and CNWM respectively, and the masses of individual organs have been adjusted to the Chinese reference data. The calculation of organ-absorbed doses and effective doses were performed with the Monte Carlo transport code MCNPX. The resulting dose conversion coefficients were compared with those published in ICRP Publication 116, which represents the reference Caucasian. The organ-absorbed dose conversion coefficients of most organs are in good agreement with the results in ICRP Publication 116, however, obvious discrepancies are observed for some organs and certain geometries. For neutrons with energies above 2 MeV, the effective dose conversion coefficients of Chinese reference adult are almost identical to those of ICRP Publication 116 in AP, PA, ROT and ISO geometries. When energies range from $10^{-8}$ MeV to 1 MeV, differences are within 10% in AP (5%), PA (8%) and ROT (-4%) geometries. However, relatively large discrepancies are shown in lateral and ISO geometries when energies are below 1 MeV, with differences of -15% for LLAT, -20% for RLAT and -12% for ISO, respectively.

## 1 Introduction

The radiation protection quantities, such as equivalent dose and effective dose, which provide a quantitative description of biological effect due to radiation exposures, play an important role in radiation protection. The equivalent dose is defined as the mean absorbed dose of the organ or tissue multiplied by the corresponding radiation weighting factor, and the effective dose can be obtained by summation of the modified equivalent dose in various organs and tissues of the human body by tissue weighting factors [1]. However, these protection quantities are unreachable through the method in the definition since organ-absorbed dose can't be measured directly. Therefore, International Commission on Radiological Protection (ICRP) Publication 74 introduces dose conversion coefficients relating protection quantities to the 'operational quantities' [2], e.g. fluence and air kerma, which can be measured directly at workplace. As a result, in the last few decades, many computational models of human body starting from simple mathematical models to voxel models and even to boundary based models have been established for calculating the various dose conversion coefficients in radiation protection.

Mathematical phantoms, which have been widely used for decades to determine radiation dose from both internal and external radiation exposures, use various mathematical equations (cones, spheres, planes, etc) to represent the various tissues, lung and skeletal regions of the body [3,4]. The reference dose conversion coefficients provided by ICRP publication 74 [2] were calculated based on the Medical Internal Radiation Dose (MIRD) phantom included in the MIRD pamphlet NO.5 by the MIRD Committee.

*Supported by National Natural Science Foundation of China (11275110, 11375103), and the National Key Scientific Instrument and Equipment Development Projects, China [number: 2012YQ180118].
1) E-mail: qiurui@tsinghua.edu.cn

With the rapid development of medical imaging technology and computers, high resolution digital images of internal anatomy become available. The first voxel model, whose data were from the computed tomography (CT) scans of a female cadaver, was reported by Gibbs et al in 1984 [5]. Since then, researchers have made considerable efforts to construct voxel models based on CT, magnetic resonance imaging (MRI) or color photographs. To date, more than 30 voxel-based models have been established, including an adult male phantom constructed by Zubal et al [6], NORMAN and NAOMI introduced by Dimbylow [7, 8], VIP-Man developed by Xu [9], 8 month pregnant woman developed by Shi and Xu [10], GSF-series voxel models developed by Zankl et al [11], the MAX and FAX developed by Kramer et al [12, 13], Otoko established by Saito et al [14], KORMAN developed by Lee et al [15], Korean-Man reported by Kim et al (2008) [16] and so on. In ICPR Publication 110, voxel phantom was decided to be the ICRP reference phantom [17]. In China, Zhang et al (2005) reported the first Chinese voxel model CNMAN, which was constructed from the Chinese Visible Human Project data set [18]. Zhang et al (2007) reported another visible Chinese human phantom (VCH) [19]. In 2009, a high-resolution voxel model of Chinese adult reference male based on a previous individual voxel model, denoted as CAM or CRAM, was constructed by Liu et al [20].

Recently, surface human phantoms, which were constructed by converting their tomographic voxel phantoms or directly developed from CT images, were introduced by several investigators. Surface phantoms make use of Polygon Meshes (PM) or Non-Uniform Rational B-spline (NURBS) to describe the boundaries of organs, tissues, skeletons and body outline. Because of its flexibility to adjustment and authenticity of the anatomy, many surface human phantoms have been developed in the past 10 years. These include the surface phantoms developed by Segars et al in 2001 [21], Xu et al in 2007 [22], Bolch et al in 2007 [23], Stabin et al in 2008 [24] and Kramer et al in 2010 [25].

Based on these anatomically realistic phantoms, broad information on operational aspects of diagnostic radiology, radionuclide therapy, nuclear medicine as well as radiation protection has been obtained. However, more attention is focused on simulations of photon transport due to its wide availability in current medical physics. Since the particle transport, nuclear interactions energy spectra and energy-loss mechanism vary a lot between types of irradiation source, it is necessary to investigate the computational dose under various circumstances. Neutron is another important but more complex source type. Neutron radiation exposures are associated with diverse human activities, such as nuclear power generation, high-energy particle acceleration and deep space explorations, where the energy spectra can range from the thermal region to above GeV levels, and always involve a variety of secondary particles. Neutron dose conversion coefficients based on ICRP reference voxel phantoms have been documented in ICRP report 116 [26]. However, there hasn't been available neutron dosimetric data based on Chinese reference voxel phantom in China.

Therefore, in present work, a new set of neutron-fluence-to-dose conversion coefficients were calculated based on the Chinese reference adult male(CRAM) and female(CRAF) voxel phantoms with energy ranging from $10^{-8}$ MeV to 20 MeV, under six idealized external neutron exposures: anterior-posterior (AP), posterior anterior (PA), left lateral (LLAT), right lateral (RLAT), rotational (ROT), and isotropic (ISO) geometries. The Monte Carlo code MCNPX was used in the calculation of organ-absorbed dose. The dose conversion coefficients from CRAM and CRAF were compared with those from ICRP Publication 116, where the phantoms were developed to represent Caucasian.

## 2 Materials and method

### 2.1 CRAM and CRAF voxel phantoms

Chinese reference adult male (CRAM) and female (CRAF) voxel phantoms, where the height, weight, and organ mass matched the Chinese reference value, were developed based on the pre-established Chinese individual voxel models CNMAN and CNWN respectively. The previous models CNMAN and CNWM were established from two set of high resolution color photographs. The set of photographs for CNMAN was from a 35 year-old Chinese male cadaver (height 170 cm, weight 65 kg), and another set of photographs for CNWM was from a 22 year-old Chinese female cadaver (height 162 cm, weight 54 kg). Liu et al described the process of constructing CRAM from the original model CNMAN in an article published previously in details [20]. Our research group previously introduced a Chinese adult female voxel phantom CNMN in 2009, using the same method as CNMAN by Zhang et al in 2007 [18]. In present work, CRAF was developed based on the original model CNWM, using the similar method as described by Liu et al in 2009 [20].

Initially, the in-plane voxel size was properly scaled in the original phantoms according to the skeleton reference volume of Asian. The requirement led to voxel dimensions in-plane of 0.613×0.613mm$^2$ for CNWM. The skeletons of CNWM were then respectively sub-segmented into 19 specific sites, each of which was subdivided into cortical bone and spongiosa. The distribution of red bone marrow, yellow bone marrow, bone trabecula, and miscellaneous tissue were treated appropriately, and their proportion matched the data of ICRP Publications 70 [27] and 89 [28] well. A more detailed description about bone can be found in reference [29].

Some important organs, including extra-thoracic, eyes, oral mucosa, lymph nodes, tooth, blood vessel and some organ walls were not segmented in CNWN, these organs were then sub-segmented. Several organs, such as eyes, lymph nodes and blood vessel, were segmented directly from the original color photographs. The organ like oral mucosa, which was hard to be segmented directly, was divided as a thin layer of oral cavity in the surface with the method of "erosion", which will descried in the following paragraph. Similarly, the segmentation of the wall from the content for organs, such as stomach, upper large intestine (ULI), lower large intestine (LLI), small intestine (SI), gall bladder (GB) and urinary bladder (UB), was also conducted using the "erosion" method, where the wall was segmented as the outer layer of the content with the proper thickness.

For the purpose of developing voxel models to represent Chinese reference female, the adjustment of organ mass was carried out by changing the organ volume, which could be derived from the reference mass divided by its corresponding density taken from ICRU publication 46 [30]. For organs whose volume lager than the respective reference value, the "grow" method was used to search and change the voxel boundary of the organ to adipose tissue until the reference volume was achieved. On the contrary, an opposite procedure known as "erosion" method, where the adjacent adipose tissue voxels were changed to organ tissue until the reference value was matched, was used for organs' volume smaller than the respective reference value. After the modulation, the organ mass in CNWM agree with the Chinese reference data. However, the whole body mass was still a little larger than the reference weight 54 kg of Chinese female. Therefore, the whole body weight adjustment was implied by reducing the adipose tissue. Eventually, the weight and height of CRAF (54kg, 160cm) was consistent with the Chinese reference value after the whole adjustment mentioned above. Fig. 1. shows the 3D view for bones and organs in CRAM and CRAF.

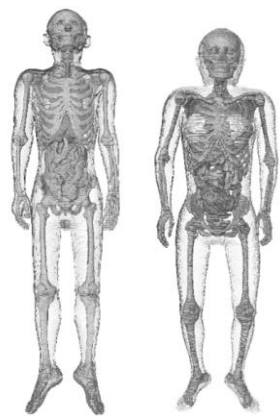

Fig. 1. 3D view for bones and organs in CRAM (left) and CRAF (right)

## 2.2 Monte Carlo simulation for CRAM and CRAF phantoms

MCNPX was used in the calculation with CRAM and CRAF phantoms to determine the organ absorbed dose from external mono-energetic neutron irradiation. As the neutron cross section is crucial to the calculation, the cross section library used in this work is shown in Table 1. The $S(\alpha,\beta)$ scattering was considered for neutrons below 4 eV, and the corresponding cross section data adopted were from those for light water at 300 K in the TMCCS library (data card: MTm: 1wtr.01t). CRAM and CRAF voxel phantoms were implemented into MCNPX by using repeated geometry structure, i.e. Universe, Lattice and Filled cards. With MATLAB programming, the voxel phantom data was converted to MCNP syntax. The tissue densities and elemental compositions used in CRAM and CRAF were obtained from ICRU publication 46 [30].

In configuration of the irradiation source, parallel neutron beams that directed perpendicularly to the longitudinal axis of the body were set for AP, PA, LLAT and RLAT geometries, respectively. A full 360º rotation of neutron beams around the longitudinal axis of the fixed body was set for ROT geometry. Neutron beams projected with no preference in direction, where the particle flux per unit solid angle remained constant, were configured for ISO geometry. Incident source energies ranging from $10^{-8}$ MeV to 20 MeV were investigated. For neutron energy below 20 MeV, the absorbed dose for tissues and organs consist of two parts, the dose contributed by heavy charged particles and recoil nucleon which were originally generated by neutron and the dose produced by the secondary photons. To record the energy deposition of neutrons and photons, the original F6 tally from MCNPX was used in this work.

Table 1. The cross-section library of neutron used for calculation in this work

| Element | Atomic number | Material identification | Cross-section library | Source | Evaluation date | Emax (MeV) | T(K) | Number of energy points |
|---|---|---|---|---|---|---|---|---|
| H  | 1  | 1001.24c  | la150n | B-VI.6 | 1998 | 150 | 293.6 | 686 |
| C  | 6  | 6000.24c  | la150n | B-VI.6 | 1996 | 150 | 293.6 | 1267 |
| N  | 7  | 7014.24c  | la150n | B-VI.6 | 1997 | 150 | 293.6 | 1824 |
| O  | 8  | 8016.24c  | la150n | B-VI.6 | 1996 | 150 | 293.6 | 1935 |
| Na | 11 | 11023.60c | endf60 | B-VI.1 | 1977 | 20  | 293.6 | 2543 |
| Mg | 12 | 12000.60c | endf60 | B-VI.0 | 1978 | 20  | 293.6 | 2525 |
| P  | 15 | 15031.24c | la150n | B-VI.6 | 1997 | 150 | 293.6 | 990 |
| S  | 16 | 16000.60c | endf60 | B-VI.0 | 1979 | 20  | 293.6 | 8382 |
| Cl | 17 | 17000.60c | endf60 | B-VI.0 | 1967 | 20  | 293.6 | 1816 |
| Ar | 18 | 18000.59c | misc5xs | LANL/T | 1982 | 20  | 293.6 | 252 |
| K  | 19 | 19000.60c | endf60 | B-VI.0 | 1974 | 20  | 293.6 | 1767 |
| Ca | 20 | 20000.24c | la150n | B-VI.6 | 1997 | 150 | 293.6 | 4470 |
| Fe | 26 | 26000.55c | rmccs  | LANL/T | 1986 | 20  | 293.6 | 6899 |
| I  | 53 | 53127.60c | endf60 | LANL/T | 1991 | 30  | 293.6 | 7888 |

**2.3 Dose conversion coefficients calculation**

Original quantities, which were directly obtained from MCNPX, were absorbed doses of tissues and organs. These quantities were then normalized by incident neutron fluence to be expressed as the fluence-to-dose conversion coefficients. However, the dose to red bone marrow (RBM) cannot be calculated directly due to its small dimensions. The tree-correction-factor (3CFs) method was used to derive the RBM dose. In the meanwhile, the dose to spongiosa was used as a surrogate for the bone surface dose. The neutron radiation weighting factor ($\omega_R$) recommended by ICRP Publication 103 was used to calculate the equivalent dose for each organ under incident levels of energy [1]. The $\omega_R$ used in this study are expressed in the Eq. (1)

$$\omega_R = \begin{cases} 2.5 + 18.2 e^{-[\ln(E_n)]^2/6}, & E_n < 1 MeV \\ 5.0 + 17.0 e^{-[\ln(E_n)]^2/6}, & 1 MeV \leq E_n \leq 50 MeV \\ 2.5 + 3.25 e^{-[\ln(0.04 E_n)]^2/6}, & E_n > 50 MeV \end{cases} \quad (1)$$

Where $E_n$ is the source energy, in unit of MeV. The whole body effective dose, acquired by multiplying the equivalent dose by the specialized tissue weighting factor ($\omega_T$) defined in ICRP Publication 103, can be calculated as Eq. (2).

$$E = \sum_T \omega_T [\frac{H_T^M + H_T^F}{2}]. \quad (2)$$

Where $H_T^M$ and $H_T^F$ represent the organ equivalent dose for male and female respectively.

**3 Result and discussion**

**3.1 Dose conversion coefficients for idealized external neutron exposures**

Once the neutron fluence $\Phi$, the organ absorbed dose $D_T$, and the whole body effective dose $E_T$ was obtained in different radiation geometries and different neutron energies, the organ absorbed dose and the whole body effective dose conversion coefficients were calculated by $D_T/\Phi$ and $E_T/\Phi$ respectively. Finally, the total amount of 3588 organ absorbed dose conversion coefficients were calculated for CRAM, corresponding to 26 organs, 6 different radiation geometries and 23 energy groups ranging from $1\times10^{-8}$ MeV to 20 MeV. Meanwhile, 4278 organ absorbed dose conversion coefficients were calculated for CRAF, corresponding to 31 organs, 6 different radiation geometries and 23 energy groups ranging from $1\times10^{-8}$ MeV to 20 MeV. In addition, 138 effective dose conversion coefficients were obtained for Chinese reference adult. The result of organ-absorbed-dose conversion coefficients for the lung and stomach in CRAM and CRAF for various idealized geometries and energies is shown in Table 2 and Table 3. The statistical uncertainties of the result in the simulation for CRAM and CRAF were less than 5%, and only 5% of all dose data are reported with uncertainties greater than 1%. Overall the precision is satisfactory

Table 2. Fluence-to-dose conversion coefficients (PGy cm$^2$) for stomach under various geometries

| Energy (MeV) | Absorbed dose conversion coefficients in various geometries(pGy cm$^2$) | | | | | | | | | | | |
|---|---|---|---|---|---|---|---|---|---|---|---|---|
| | AP | PA | LLAT | RLAT | ROT | ISO | AP | PA | LLAT | RLAT | ROT | ISO |
| stomach | | | CRAM | | | | | | CRAF | | | |
| 1.00E-08 | 1.46 | 0.59 | 0.85 | 0.27 | 0.83 | 0.7 | 1.33 | 0.65 | 0.67 | 0.25 | 0.75 | 0.70 |
| 1.00E-07 | 2.21 | 0.92 | 1.32 | 0.43 | 1.25 | 0.99 | 2.01 | 1.03 | 1.03 | 0.39 | 1.16 | 0.97 |
| 1.00E-06 | 3.12 | 1.38 | 1.97 | 0.62 | 1.82 | 1.4 | 2.82 | 1.53 | 1.54 | 0.54 | 1.72 | 1.33 |
| 1.00E-05 | 3.49 | 1.59 | 2.24 | 0.72 | 2.04 | 1.6 | 3.26 | 1.77 | 1.78 | 0.64 | 1.99 | 1.60 |
| 1.00E-04 | 3.52 | 1.62 | 2.28 | 0.74 | 2.08 | 1.64 | 3.38 | 1.8 | 1.81 | 0.67 | 2.02 | 1.61 |
| 1.00E-03 | 3.45 | 1.67 | 2.24 | 0.75 | 2.06 | 1.6 | 3.27 | 1.86 | 1.83 | 0.67 | 2.06 | 1.64 |
| 5.00E-03 | 3.40 | 1.67 | 2.21 | 0.76 | 2.01 | 1.59 | 3.26 | 1.9 | 1.82 | 0.67 | 2.07 | 1.62 |
| 1.00E-02 | 3.44 | 1.69 | 2.20 | 0.75 | 2.08 | 1.6 | 3.25 | 1.88 | 1.85 | 0.69 | 2.06 | 1.59 |
| 5.00E-02 | 3.78 | 1.81 | 2.37 | 0.81 | 2.22 | 1.69 | 3.53 | 2.04 | 1.95 | 0.73 | 2.2 | 1.67 |
| 1.00E-01 | 4.21 | 1.92 | 2.67 | 0.88 | 2.44 | 1.84 | 3.87 | 2.16 | 2.17 | 0.80 | 2.33 | 1.78 |
| 2.00E-01 | 5.22 | 2.10 | 3.36 | 1.03 | 2.91 | 2.18 | 4.63 | 2.47 | 2.49 | 0.95 | 2.77 | 2.03 |
| 5.00E-01 | 8.55 | 2.76 | 6.04 | 1.50 | 4.61 | 3.43 | 7.36 | 3.32 | 4.43 | 1.38 | 4.24 | 3.09 |
| 1.00E+00 | 12.19 | 3.29 | 9.18 | 2.03 | 6.60 | 4.98 | 10.52 | 4.18 | 6.82 | 1.93 | 6.00 | 4.40 |
| 2.00E+00 | 22.51 | 8.82 | 19.97 | 6.50 | 13.74 | 10.63 | 20.71 | 10.51 | 16.31 | 5.91 | 13.39 | 9.95 |
| 4.00E+00 | 36.05 | 17.93 | 33.55 | 13.89 | 23.89 | 18.93 | 33.49 | 19.95 | 28.6 | 12.46 | 23.6 | 18.09 |
| 6.00E+00 | 44.57 | 26.06 | 42.66 | 21.73 | 32.26 | 26.12 | 42.14 | 28.24 | 38.01 | 19.27 | 31.81 | 25.47 |
| 8.00E+00 | 50.25 | 32.78 | 48.96 | 28.37 | 38.08 | 31.65 | 48.28 | 34.76 | 44.22 | 25.55 | 38.04 | 31.36 |
| 1.00E+01 | 56.26 | 37.62 | 54.88 | 32.72 | 43.32 | 36.13 | 54.31 | 39.35 | 50.19 | 29.47 | 43.22 | 35.77 |
| 1.20E+01 | 61.81 | 42.44 | 60.31 | 36.76 | 47.76 | 40.32 | 59.43 | 44.42 | 55.2 | 33.03 | 48.03 | 39.70 |
| 1.40E+01 | 64.88 | 46.89 | 63.87 | 41.63 | 51.68 | 43.77 | 62.83 | 48.65 | 59.05 | 37.22 | 52.18 | 43.72 |
| 1.60E+01 | 67.66 | 50.83 | 67.12 | 45.32 | 55.07 | 47.12 | 65.89 | 52.24 | 62.67 | 40.84 | 55.36 | 47.01 |
| 1.80E+01 | 69.22 | 53.79 | 69.24 | 48.79 | 57.26 | 49.56 | 67.35 | 55.33 | 64.72 | 43.89 | 57.68 | 49.51 |
| 2.00E+01 | 72.91 | 57.46 | 73.75 | 52.47 | 61.52 | 53.02 | 71.65 | 59.48 | 68.97 | 47.56 | 61.68 | 52.97 |

Table 3. Fluence-to-dose conversion coefficients (PGy cm2) for lung under various geometries

| Energy (MeV) | Absorbed dose conversion coefficients in various geometries(pGy cm$^2$) | | | | | | | | | | | |
|---|---|---|---|---|---|---|---|---|---|---|---|---|
| | AP | PA | LLAT | RLAT | ROT | ISO | AP | PA | LLAT | RLAT | ROT | ISO |
| Lung | | | CRAM | | | | | | CRAF | | | |
| 1.00E-08 | 1.21 | 0.95 | 0.50 | 0.49 | 0.84 | 0.74 | 0.89 | 1.11 | 0.37 | 0.38 | 0.75 | 0.69 |
| 1.00E-07 | 1.82 | 1.52 | 0.74 | 0.73 | 1.25 | 1.03 | 1.3 | 1.71 | 0.53 | 0.56 | 1.12 | 0.96 |
| 1.00E-06 | 2.57 | 2.27 | 1.04 | 1.03 | 1.8 | 1.46 | 1.9 | 2.55 | 0.76 | 0.79 | 1.6 | 1.34 |
| 1.00E-05 | 2.89 | 2.62 | 1.16 | 1.16 | 2.00 | 1.67 | 2.19 | 2.91 | 0.88 | 0.91 | 1.86 | 1.56 |
| 1.00E-04 | 2.89 | 2.65 | 1.16 | 1.15 | 2.04 | 1.68 | 2.26 | 2.94 | 0.9 | 0.95 | 1.9 | 1.62 |
| 1.00E-03 | 2.81 | 2.62 | 1.12 | 1.12 | 2.01 | 1.65 | 2.28 | 2.89 | 0.91 | 0.95 | 1.89 | 1.62 |
| 5.00E-03 | 2.75 | 2.6 | 1.11 | 1.1 | 1.97 | 1.62 | 2.24 | 2.85 | 0.9 | 0.94 | 1.88 | 1.58 |
| 1.00E-02 | 2.76 | 2.59 | 1.11 | 1.1 | 1.98 | 1.61 | 2.25 | 2.85 | 0.91 | 0.95 | 1.89 | 1.57 |
| 5.00E-02 | 2.96 | 2.74 | 1.21 | 1.18 | 2.08 | 1.66 | 2.42 | 3.1 | 0.96 | 1.00 | 1.97 | 1.66 |
| 1.00E-01 | 3.35 | 2.97 | 1.38 | 1.34 | 2.33 | 1.83 | 2.6 | 3.44 | 1.03 | 1.06 | 2.14 | 1.74 |
| 2.00E-01 | 4.35 | 3.58 | 1.77 | 1.72 | 2.87 | 2.26 | 2.96 | 4.38 | 1.16 | 1.23 | 2.52 | 2.03 |
| 5.00E-01 | 7.77 | 6.03 | 3.07 | 2.99 | 4.94 | 3.77 | 4.56 | 7.71 | 1.72 | 1.84 | 4.15 | 3.23 |
| 1.00E+00 | 11.55 | 8.82 | 4.57 | 4.43 | 7.36 | 5.81 | 6.66 | 11.53 | 2.45 | 2.68 | 6.21 | 5.00 |
| 2.00E+00 | 22.42 | 19.67 | 10.13 | 9.9 | 15.63 | 12.68 | 15.59 | 23.23 | 6.52 | 7.11 | 14.02 | 11.56 |
| 4.00E+00 | 35.73 | 32.96 | 18.24 | 17.77 | 26.51 | 22.03 | 27.32 | 37.12 | 13.06 | 14.03 | 24.6 | 20.78 |
| 6.00E+00 | 43.91 | 41.53 | 25.49 | 24.92 | 34.62 | 29.55 | 36.01 | 45.71 | 19.96 | 20.96 | 32.87 | 28.49 |
| 8.00E+00 | 49.54 | 47.8 | 31.15 | 30.41 | 40.35 | 35.02 | 42.25 | 51.51 | 25.35 | 26.36 | 38.72 | 34.10 |
| 1.00E+01 | 55.5 | 53.66 | 35.75 | 35.03 | 45.66 | 39.83 | 47.83 | 57.78 | 29.64 | 30.62 | 44.16 | 39.02 |
| 1.20E+01 | 60.73 | 59.1 | 39.48 | 38.77 | 50.29 | 43.91 | 52.7 | 63.31 | 32.99 | 34.09 | 48.68 | 43.12 |
| 1.40E+01 | 63.7 | 62.54 | 43.17 | 42.45 | 53.86 | 47.36 | 56.42 | 66.45 | 36.84 | 37.78 | 52.33 | 46.83 |
| 1.60E+01 | 66.7 | 66.05 | 46.42 | 45.59 | 57.12 | 50.58 | 59.91 | 69.62 | 39.99 | 40.91 | 55.56 | 50.07 |
| 1.80E+01 | 68.33 | 68.04 | 49.01 | 48.2 | 59.21 | 52.81 | 62.21 | 71.12 | 42.69 | 43.46 | 57.92 | 52.43 |
| 2.00E+01 | 72.86 | 72.55 | 52.84 | 52.06 | 63.44 | 56.73 | 66.25 | 75.59 | 46.02 | 46.88 | 62.09 | 56.53 |

## 3.2 Neutron Dose conversion coefficients at different energies

The shape of the curves showing the fluence-to-absorbed dose conversion coefficients depending on neutron energy are more or less similar for all the organs and geometries, which is in accordance with the varying predominance of types of particle interactions. Within the thermal regime, most of the direct energy deposition is contributed by the gamma ray through neutron-capture reaction. When the neutron energy increase from $10^{-8}$ MeV to $10^{-6}$ MeV, a slight increment can be observed for the dose conversion coefficient. At energy ranging from 1eV to 10keV, inelastic scattering and resonant capture reaction take place, which lead to a flat region displayed. With the energy exceeding 1 MeV, sharp increases are exhibited due to large energy transferring during elastic scattering. Anatomical variation among different models is comparatively more important in low-energy irradiation condition since the particle deposit most of their energy around the sites where they enter the human body.

## 3.3 Comparison of conversion coefficients among different irradiation conditions and different models

Absorbed dose conversion coefficients for stomach in CRAM, for liver in CRAM, for Lung in CRAF, and for red bone marrow (RBM) in CRAF in six idealized irradiation geometries are shown in Fig. 2(a), (b), (c), and (d), respectively. Since stomach and liver are asymmetry organs located superficially in the left and right region of body trunk respectively, dose coefficients for these two organs irradiated from reverse lateral direction, i.e. RLAT for stomach and LLAT for the liver, are least among the six irradiation geometries, and dose coefficients of AP irradiation geometry of stomach and liver are the maximum. Correspondingly, for organs like lung and red bone narrow which are located symmetrically in both sides of human body, the dose conversion coefficients are almost the same in LLAT and RLAT geometries.

Dose coefficients from CRAM and CRAF are compared with those from ICRP based voxel phantoms. For most organs, dose coefficients from both the Chinese reference phantoms and ICRP based phantoms are consistent. Fig. 3(a) and (b) shows the fluence-to-absorbed dose conversion coefficients for the stomach in RLAT and AP irradiation geometries, respectively. In comparison with the ICRP based male and female voxel models, dose coefficients for stomach from CRAM and CRAF are obviously higher in RLAT geometry but well agreement in AP geometry. This can be understood by the fact that Caucasians have higher height and heavier weight than Chinese, which lead to larger lateral shielding thickness in Caucasian models than that in Chinese models. However, the variation of the anterior side thickness of the different models is less pronounced than that of the lateral side since thick layer of fat tissue can be found in all of models. As a result, only very small discrepancies are noticeable for stomach in AP geometries for different models. Dose conversion coefficients for bladder of different model in PA and AP irradiation geometries are shown in Fig. 3(c) and (d), respectively. Similarly, since there is a considerable amount of muscle and adipose tissue concentrating the back of the ICRP reference based voxel models, dose coefficients from PA of Chinese reference voxel models are higher when compared with the ICRP models. It can be also observed that dose coefficients for bladder of all models agree well in AP geometry also for the reason described above.

Considering that the ISO and ROT geometries are to some extent the combination of other four idealized geometries, smaller difference is usually observed for most organs in these two geometries between Chinese reference based voxel models and the ICRP voxel models. Fig. 4 (a), (b), (c) and (d) show the dose coefficients of different models for lung in ISO, esophagus in ROT, liver in ROT and thyroid in ISO, respectively. It can be obviously observed that dose coefficients for lung, esophagus and liver are basically the same in different models. However, relatively large discrepancy for thyroid in different models due to its small size.

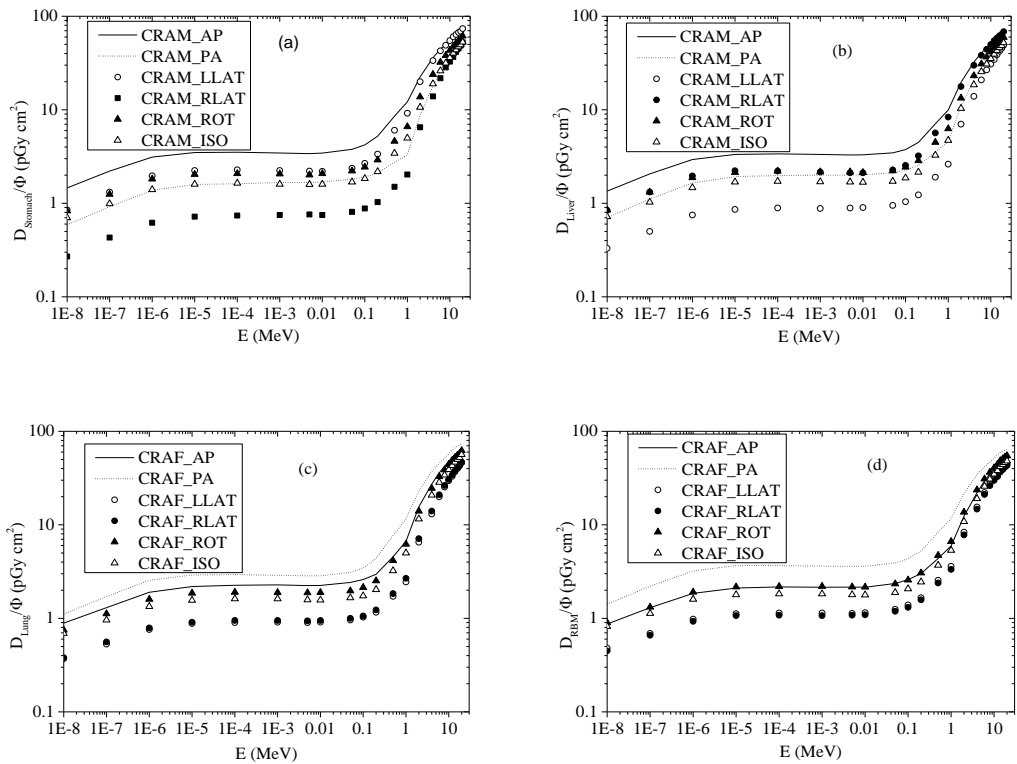

Fig. 2. Comparison of absorbed dose per fluence for the stomach of the CRAM (a), liver of CRAM (b), lung of CRAF (c), RBM of CRAF (d) in six ideal radiation geometries.

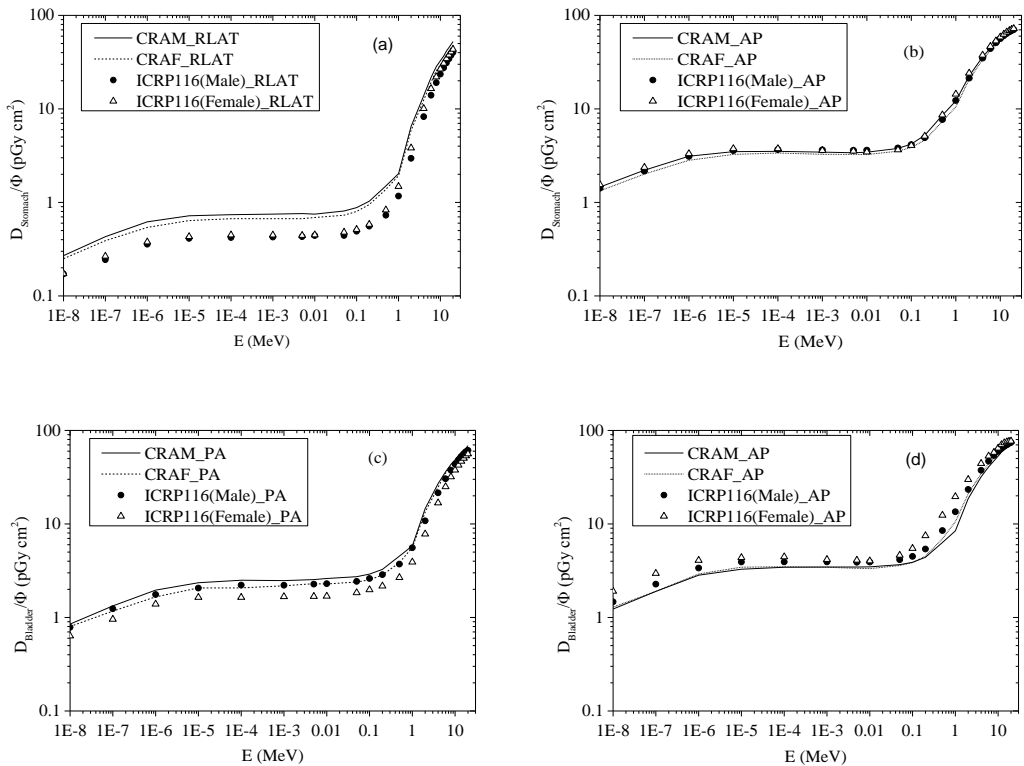

Fig. 3. Comparison of organ absorbed dose conversion coefficients from Chinese reference adult phantoms with values obtained from the ICRP 116 for stomach in RLAT (a) and AP (b), bladder in PA (c) and AP (d).

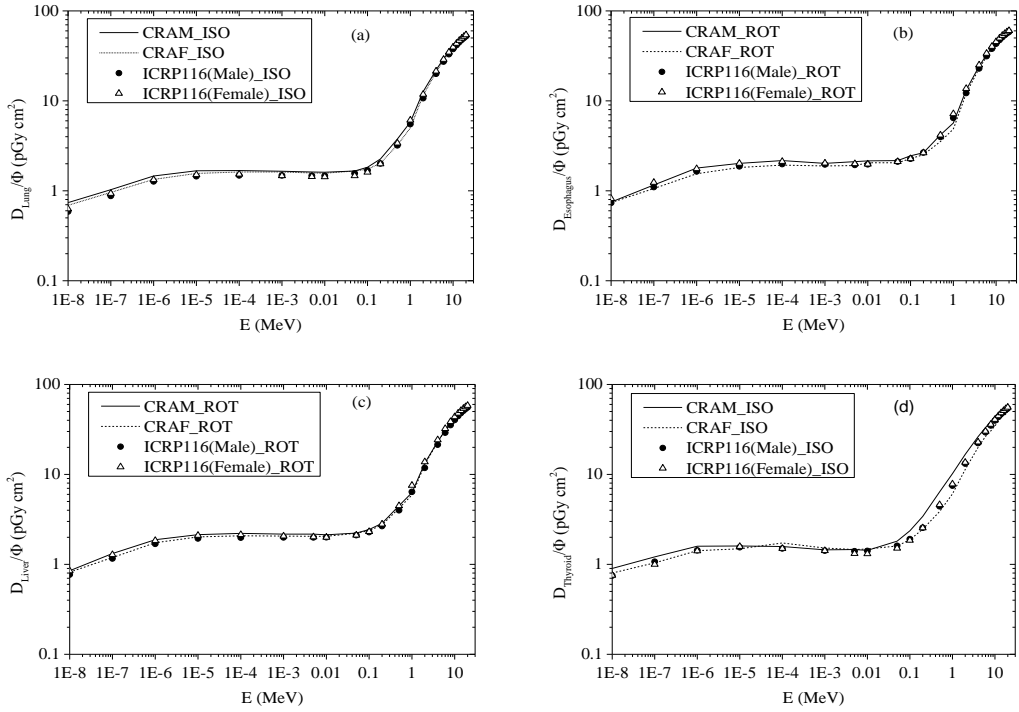

Fig. 4. Comparison of organ absorbed dose conversion coefficients among CRAM, CRAF, ICRP male and female phantom for lung in ISO (a), esophagus in ROT (b), liver in ROT (c) and thyroid in ISO (d).

The organ absorbed dose conversion coefficients of Chinese adult reference phantoms CRAM and CRAF have been used to calculate the effective dose conversion coefficients. The radiation weighting factor of neutron and the organ-specific tissue weighting factors from ICRP 103 were applied to calculated the effective doses according to the Equation (2) introduced also by ICRP 103. Fig. 5 (a) shows the effective dose conversion coefficients of the Chinese reference voxel phantoms under six idealized geometries. As is shown in the figure, the effective dose yielded by the AP geometry is the largest among the six irradiation condition, since most organs which are critical to the effective dose are located within the frontal portion of the body. However, the arms and increased body thickness along the coronal axis provide long distance for particle to transport to the important organs, as a result of which, RLAT and LLAT report lower effective doses. Considering the particle in ROT and ISO geometries are uniformly emitted, the effective doses from ROT and ISO are in the middle status.

Ratios of the effective dose conversion coefficients from ICRP 116 to those from Chinese adult reference phantoms under six idealized geometries are shown in Fig. 5 (b). For energies above 2 MeV, the effective dose coefficients of ICRP 116 is almost the same as those of Chinese reference adult phantoms in AP, PA, ROT and ISO geometries, where the difference is within 5%. Compared with the Chinese reference adult phantoms, effective doses from lateral geometries of ICRP 116 is generally 10% lower for energies above 2 MeV, in which 15% at 4 MeV and 10% at 14 MeV in RLAT geometry. At energies below 1 MeV, the difference is quite modest in AP (5%), PA (8%) and ROT (-4%) geometries. However, a relatively large discrepancy is shown in lateral and ISO geometries at energies blew 1 MeV, where the difference are -15% for LLAT, -20% for RLAT and -12% for ISO, respectively.

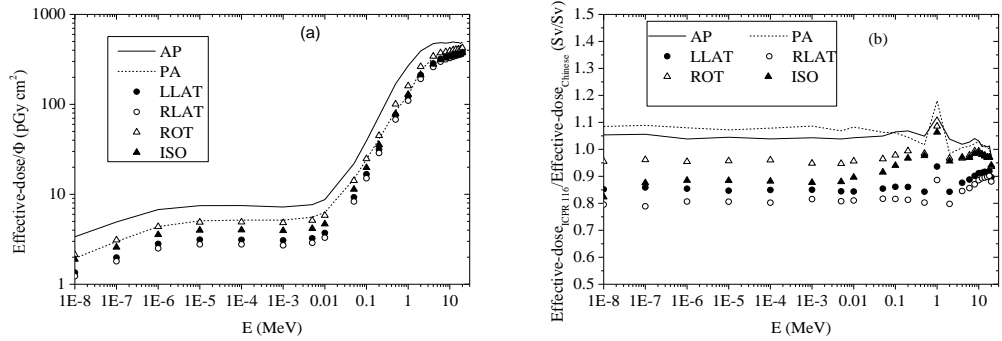

Fig. 5. The effective dose conversion coefficients of the Chinese Reference voxel phantoms under six idealized geometries (a), and ratios of the effective dose conversion coefficients from ICRP 116 data to those from Chinese adult reference phantoms under six idealized geometries (b).

## 4 Conclusion

Two phantoms CRAM and CRAF, which represent the Chinese reference adult male and female adjusted from the original phantoms CNMAN and CNWM respectively, have been utilized to calculate the neutron organ absorbed and effective dose conversion coefficients. The calculation was performed using the Monte Carlo code MCNPX under six ideal radiation geometries with energy ranging from $10^{-8}$ MeV to 20 MeV. The CRAM phantom has a height of 170 cm and a weight of 60 kg with 80 organs and tissues, and the CRAF phantom has a height of 160 cm and a weight of 54 kg with 85 organs and tissues. The organ dose conversion coefficients of CRAM and CRAF for six standard external exposures were compared with each other, where the deviation determined by the location of the organs. Considerable differences were observed for several organs under some specific irradiation geometries. This new set of dose conversion coefficients which represent the reference Chinese adult was compared with the data recommended in ICRP 116 representing the reference Caucasian. It was observed that the two sets of organ absorbed dose conversion coefficients agreed well for AP, ROT and ISO exposures. However, considerable deviations were found in lateral since the height and weight of Chinese are smaller than that of the Caucasian. As a result of an abundant amount of muscle and adipose tissue concentrating the back of the ICRP reference based voxel models, enormous difference were observed in PA irradiation geometry. The effective dose conversion coefficients of Chinese reference adult is almost identical to those of ICRP 116 for neutron energy above 2 MeV in AP, PA, ROT and ISO geometries. At energies from $10^{-8}$ MeV to 1 MeV, the difference is quite slight in AP (5%), PA (8%) and ROT (-4%) geometries. However, a relatively large discrepancy is shown in lateral and ISO geometries at energies below 1 MeV, where the difference are -15% for LLAT, -20% for RLAT and -12 for ISO, respectively.